\documentclass
	[a4paper,twocolumn,twoside,10pt,showkeys]
	{revtex4}

\usepackage{amsmath,amssymb}	
\usepackage{bm}					
\usepackage{color}					
\usepackage{graphicx}				
\usepackage{dcolumn}				

\makeatletter
\def\frontmatter@abstractwidth{0.9\textwidth}	
\makeatother

\begin{document}

\newcommand{\By}{$\times$}
\newcommand{\SqrtBy}[2]{$\sqrt{#1}$\kern0.2ex$\times$\kern-0.2ex$\sqrt{#2}$}
\newcommand{\Degree}{$^\circ$}
\newcommand{\DegreeC}{$^\circ$C}
\newcommand{\Ohmcm}{$\Omega\cdot$cm}

\title{
  First-principles Study of Rashba Spin Splitting at Strained SrTiO$_3$(001) Surfaces
}


\author{Naoya Yamaguchi}
\email[Corresponding author: ]{n-yamaguchi@cphys.s.kanazawa-u.ac.jp}
\affiliation{%
  Division of Mathematical and Physical Sciences, Graduate School of Natural Science and Technology, Kanazawa University,
Kakuma-machi, Kanazawa 920-1192, Japan
}

\author{Fumiyuki Ishii}
\email[Corresponding author: ]{ishii@cphys.s.kanazawa-u.ac.jp}
\affiliation{%
  Faculty of Mathematics and Physics, Institute of Science and Engineering, Kanazawa University,
Kakuma-machi, Kanazawa 920-1192, Japan
}

\begin{abstract}

  \vspace*{1mm}

  We investigated the Rashba spin splittings in a compressive-strained SrTiO$_3$(001) ultra thin-film using first-principles calculations.
  The effect of the polarization due to the compressive strain leads to the 2DEG with large Rashba spin splittings, where the sheet carrier density is of the same order of magnitude as that of the heterostructure LaAlO$_3$/SrTiO$_3$ ($\sim10^{14}$ cm$^{-2}$).
  Some localized surface states (SSs) show the giant Rashba coefficient $\alpha_R$ larger than 100 meV$\cdot$\AA.
\end{abstract}

\keywords{
  Density functional calculations; Single crystal surfaces; Metallic surfaces; Rashba spin splitting;
  }

\maketitle
\newpage


\section{Introduction}

Spin-to-charge conversion enables spin current to change charge current, and its phenomenon owing to the Rashba effect \citep{Bychkov_Properties_1984}, that is, the inverse Rashba-Edelstein effect (IREE) \citep{Edelstein_Spin_1990, Snchez_Spin_2013} was observed at the n-type interface in the heterostructure LaAlO$_3$/SrTiO$_3$ \citep{Lesne_Highly_2016}.
In a two-dimensional electronic system with the spatial inversion symmetry breaking, the Rashba effect occurs and induces spin splittings.
As the Rashba coefficient $\alpha_R$, the strength of the Rashba effect, becomes larger, the efficiency of the IREE may become higher \citep{Edelstein_Spin_1990}.
In our previous study for the LaAlO$_3$/SrTiO$_3$, we found the possibility of large Rashba spin splittings at the oxide surfaces or interfaces due to the strain effects which may control the polarity \citep{Yamaguchi_Strain_2017}.

The bulk SrTiO$_3$ has the strain-induced polarization, through which the strain can control the polarity \citep{Haeni_Room_2004}, and therefore an SrTiO$_3$ surface with the strain-induced polarization is a promising candidate for large Rashba spin splittings.
Actually, for the SrTiO$_3$(001) surface, the Rashba spin splitting is so small that the $k$-cubic term is more dominant than $k$-linear term \citep{Nakamura_Experimental_2012}, while there is room for discussion about large spin splittings \citep{Santander-Syro_Giant_2014, Walker_Absence_2016}.
It was reported that two-dimensional electron gas (2DEG) was induced due to the hydrogen absorption into the SrTiO$_3$(001) surface \citep{DAngelo_Hydrogen_2012, Yukawa_Electronic_2013}.
Since 2DEG can appear at ferroelectric surfaces such as a BaTiO$_3$ surface \citep{Meyerheim_BaTiO3_2012, Chen_Effects_2015}, a compressive-strained SrTiO$_3$(001) surface is also expected to have 2DEG as well as larger Rashba spin splittings.

In this paper, we investigated the 2DEG and Rashba spin splittings in a compressive-strained SrTiO$_3$(001) ultra thin-film using first-principles calculations.
The effect of the polarization due to the compressive strain induces the 2DEG with large Rashba spin splittings.
Some localized surface states (SSs) can lead to the giant Rashba coefficient $\alpha_R$ larger than 100 meV$\cdot$\AA.

\section{Computational}
Our calculations were done by using OpenMX code\citep{OpenMX}, which implements density functional theory within the generalized gradient approximation (GGA) with PBE exchange correlation functional \citep{Perdew_Generalized_1996}.
We adopted norm-conserving pseudopotentials with an energy cutoff of 300 Ry for charge density, including the 4\textit{s}, 4\textit{p}, and 5\textit{s} for Sr; 3\textit{s}, 3\textit{p}, 3\textit{d}, and 4\textit{s} for Ti; 2\textit{s} and 2\textit{p} for O.
In order to consider the spin splitting, the spin-orbit interaction was taken into account by a treatment of fully relativistic total angular momentum dependent pseudopotentials \citep{Theurich_Self_2001}.
We used a $24\times24\times1$ regular \textit{k}-point mesh.
The numerical pseudo atomic orbitals are as follows: the numbers of the \textit{s}-, \textit{p}-, and \textit{d}-character orbitals are 3, 3, and 2, respectively, for Sr; 3, 3, and 2, respectively, for Ti; 3, 3, and 1, respectively, for O.
We also used the effective screening medium (ESM) method to eliminate the dipole-dipole interaction between slab models \citep{Otani_First_2006}.

Our computational models are slab models of a superlattice (SrTiO$_3$)$_{n+0.5}$ with homogeneous SrO or TiO$_2$ terminated surfaces an experimental lattice constant of $a_0$ = 3.905\AA~\citep{Inaguma_Quantum_1992}.
We assumed the compressive strain $(a_0-a)/a$, where $a$ denotes the in-plane lattice constant.
The initial atomic configurations before the structural optimization are based on those of bulk SrTiO$_3$ with the compressive strain inducing the polarization.

\section{Results and Discussion}
We found the stable atomic configuration with the metallic SSs, and confirmed that the 2DEG emerges when the polarization breaks the spatial inversion symmetry along the surface-perpendicular direction ($z$-axis, hereafter), that is, the polarization is nonzero.
The effect of the polarization induces the 2DEG through the charge transfer, and actually it was reported that BaTiO$_3$(001) surface has 2DEG by the similar mechanism \citep{Meyerheim_BaTiO3_2012, Chen_Effects_2015}.

First, for the SrO terminations, we investigated the case where metallic SSs appear.
As shown in Fig. 1, as the number of layers in the model (or formula units $n$) or strain increases, the metallic SSs tend to be appear.
There are differences between the symmetry of the system in real space with and without metallic SS: The symmetric atomic configurations and atom-projected density of states (ADOS), or asymmetric ones.
By analyzing the ADOS shown in Fig. 2, we confirmed that in the asymmetric case there are metallic SSs (Fig. 2(a)), while only insulating states exist in the symmetric case (Fig. 2(b)).
The metallic SSs mainly consist of O-2\textit{p} bands at one surface (p-type) and Ti-3\textit{d} bands at the other (n-type).
From these results, metallic SSs are expected to emerge at the strained SrTiO$_3$(001) surface, and the thicker computational model is necessary for considering the metallic SSs.
Therefore, we focused on some cases of $n$ = 10 with metallic SSs.
For the TiO$_2$ termination as well as the SrO termination, the metallic SSs are induced.
For the compressive strain of 5\%, 6\%, and 7\%, the sheet carrier densities $n_s$ of the 2DEG are $3.42\times10^{14}$ cm$^{-2}$, $5.16\times10^{14}$ cm$^{-2}$, and $7.34\times10^{14}$ cm$^{-2}$, respectively (SrO termination); $5.05\times10^{14}$ cm$^{-2}$, $5.54\times10^{14}$ cm$^{-2}$, and $5.88\times10^{14}$ cm$^{-2}$, respectively (TiO$_2$ termination).
These values are of the same order of magnitude as that of the heterostructure LaAlO$_3$/SrTiO$_3$ ($\sim10^{14}$ cm$^{-2}$) \citep{Huijben_Electronically_2006}.
Moreover, we estimated the Rashba coefficients $\alpha_R$ near the Fermi level using an expression $\alpha_R=2E_R/k_R$, where $E_R$, $k_R$ are the Rashba energy, and the Rashba momentum offset, respectively, being obtained from the band structure inducing the Rashba spin splitting \citep{Manchon_New_2015}.
The maximum Rashba coefficients $\alpha_R^\textrm{max}$ for the case with the compressive strain of 5\%, 6\%, and 7\% are 38.9 meV$\cdot$\AA, 28.7 meV$\cdot$\AA, and 155.7 meV$\cdot$\AA~(SrO termination), respectively; 79.2 meV$\cdot$\AA, 48.7 meV$\cdot$\AA, and 21.0 meV$\cdot$\AA~(TiO$_2$ termination), respectively.
In fact, it was reported that epitaxial growth of SrTiO$_3$ films was achieved on MgO with a lattice mismatch of 7.9\% \citep{McMitchell_Two_2009}.
The metallic SSs of strained SrTiO$_3$ may induce the giant Rashba spin splitting with the $\alpha_R$ larger than 100 meV$\cdot$\AA.

Next, we investigated the SSs inducing the giant Rashba spin splitting.
Figure 3(a) shows the band structure projected to the p-type surface for the case with $n$ = 10 and the compressive strain of 7\%, and there are some spin splittings, and we focused on the notable surface Rashba states (SRS).
SRSs show the giant Rashba spin splitting of $\alpha_R$ = 155.7 meV$\cdot$\AA, and is strongly localized around the topmost oxygen atom at the surface as shown in Fig. 3(b).
Our previous study about Bi/\textit{M} surface alloys (\textit{M} = Cu, Ag, Au, Fe, Co, Ni) implied that the localization of SSs may enhance $\alpha_R$ \citep{Yamaguchi_First_2017}.
A theoretical work \citep{Nagano_A_2009} suggested an expression for $\alpha_R$: $\alpha_R=(2/c^2)\int d\vec{r}\partial_zV|\psi_{SS}|^2$, where $c$ is the velocity of light, $V$ is the potential originating from both atoms (symmetric) and the surface, and $|\psi_{SS}|^2$ is the charge density of SSs.
The calculated $|\psi_{SS}|^2$ have large amplitude at an oxygen atom.
Therefore, symmetric atomic potential and assymetric charge density give finite $\alpha_R$.
The electrons in more localized SSs feel a stronger electric field on atomic sites, which may make $\alpha_R$ larger.

\begin{figure}
  \centering
  \includegraphics[width=\columnwidth]{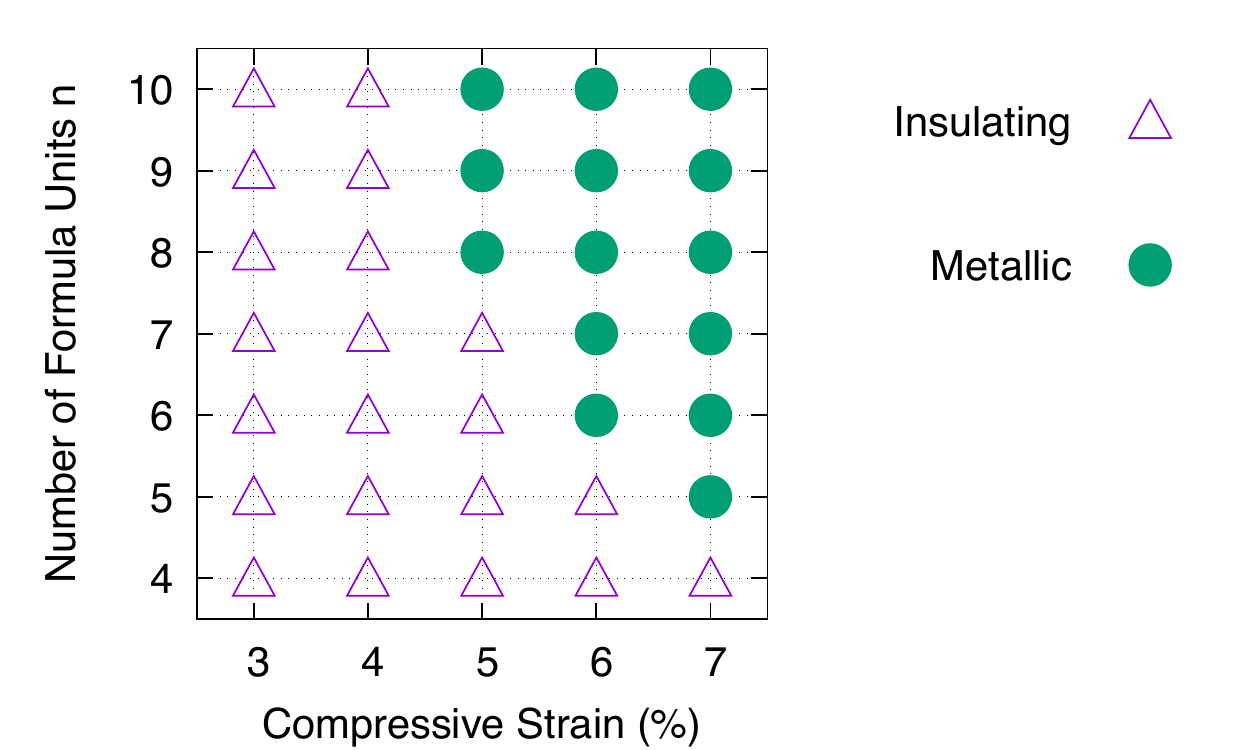}
  \caption{\label{fig:figure1}
  Thickness-strain phase diagram in terms of the emergence of the 2DEG in the case of SrO terminations. $n$ is the number of formula units. The solid circle denotes that metallic SSs exist (asymmetric atomic configurations), while the hollow triangle denotes no metallic SSs (asymmetric).}
\end{figure}

\begin{figure}
  \begin{minipage}[t]{0.49\textwidth}
    \centering
    \begin{enumerate}
	\def\labelenumi{(\alph{enumi})}
	\setcounter{enumi}{0}
      \item
    \end{enumerate}
    \includegraphics[width=\columnwidth]{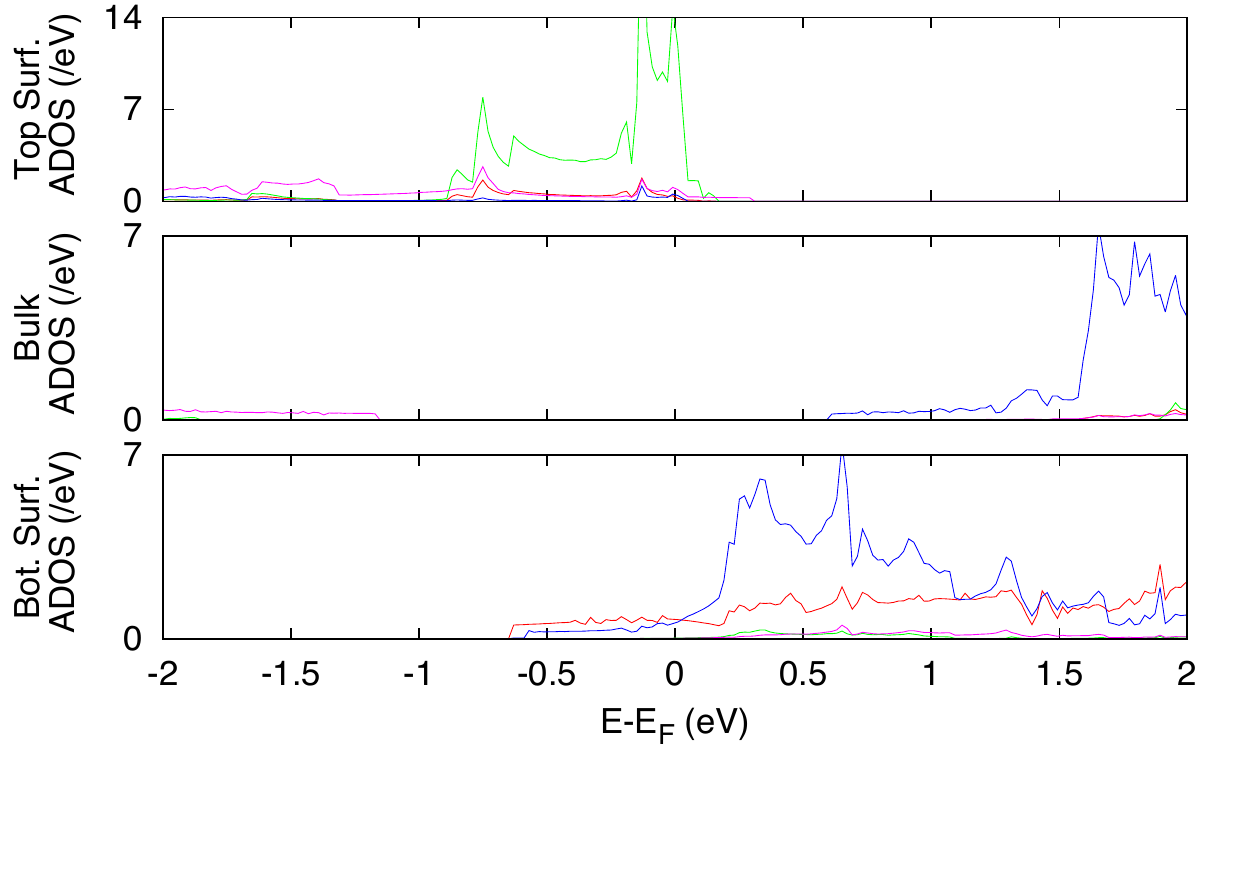}
  \end{minipage}
  \begin{minipage}[t]{0.49\textwidth}
    \centering
    \begin{enumerate}
	\def\labelenumi{(\alph{enumi})}
	\setcounter{enumi}{1}
      \item
    \end{enumerate}
    \includegraphics[width=\columnwidth]{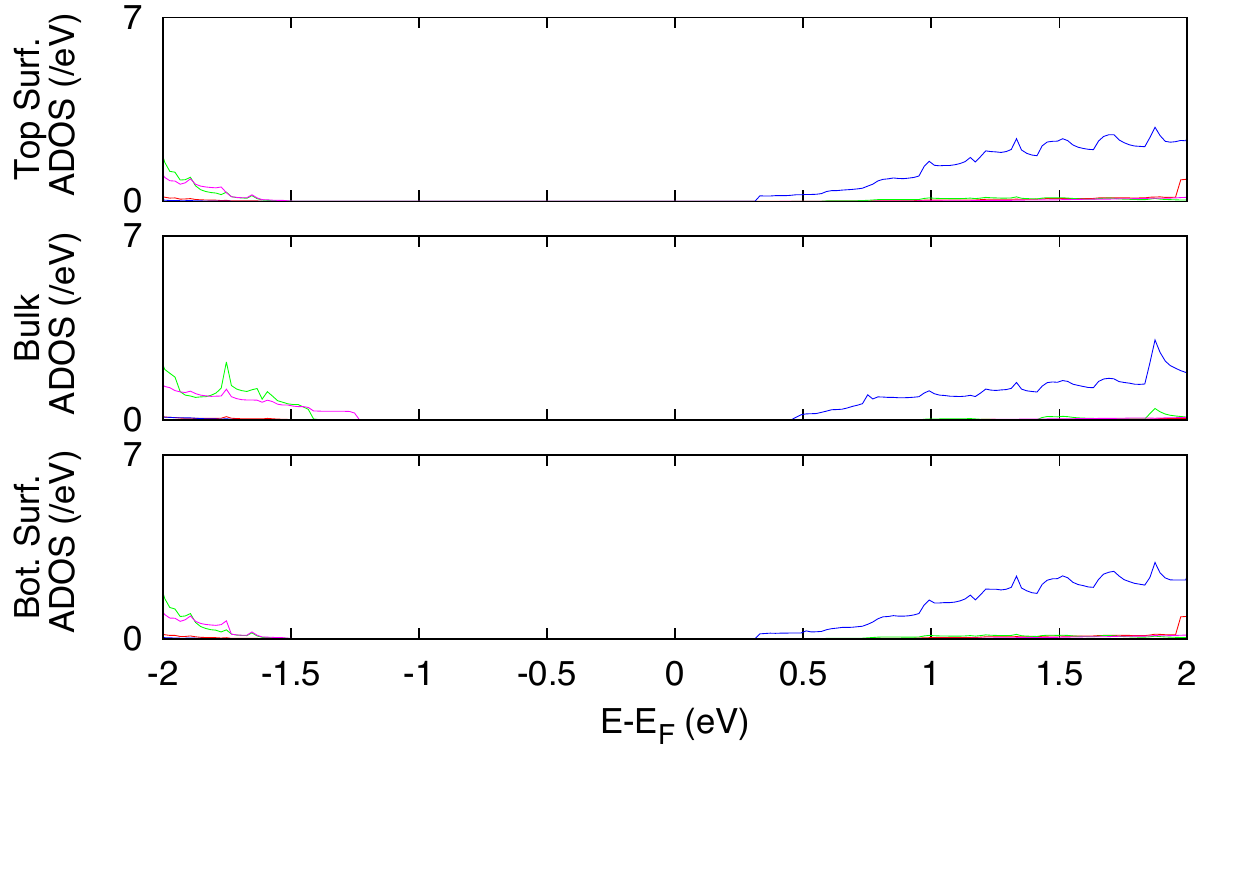}
  \end{minipage}
  \caption{\label{fig:figure2}
  Atom-projected density of states for each region (surfaces or bulk) in two cases in terms of metallic SSs in the case of SrO terminations: (a) The metallic SSs ($n$ = 10, with the compressive strain of 7\%); (b) The insulating SSs ($n$ = 10, 3\%). The origin in energy is set to the Fermi level. For the oxygen atoms, there are two cases: In the SrO or TiO$_2$ layer. The red, green, blue, and light red solid lines show the ADOS for Sr, O (in the SrO layer), Ti, and O (in the TiO$_2$ layer), respectively.}
\end{figure}

\begin{figure}
  \begin{minipage}[t]{0.49\textwidth}
    \centering
    \begin{enumerate}
	\def\labelenumi{(\alph{enumi})}
	\setcounter{enumi}{0}
      \item
    \end{enumerate}
    \includegraphics[width=\columnwidth]{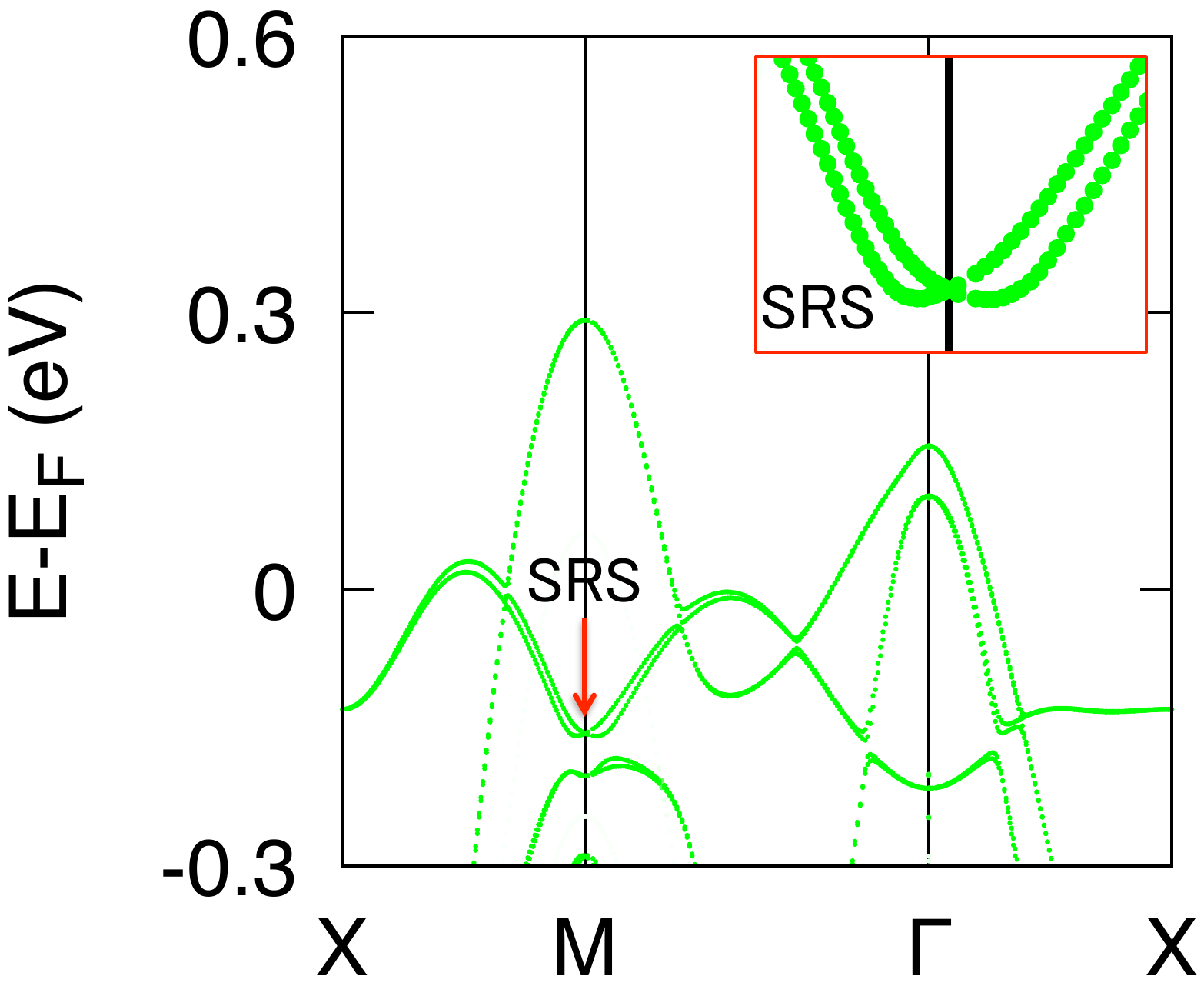}
  \end{minipage}
  \begin{minipage}[t]{0.49\textwidth}
    \centering
    \begin{enumerate}
	\def\labelenumi{(\alph{enumi})}
	\setcounter{enumi}{1}
      \item
    \end{enumerate}
    \includegraphics[width=\columnwidth]{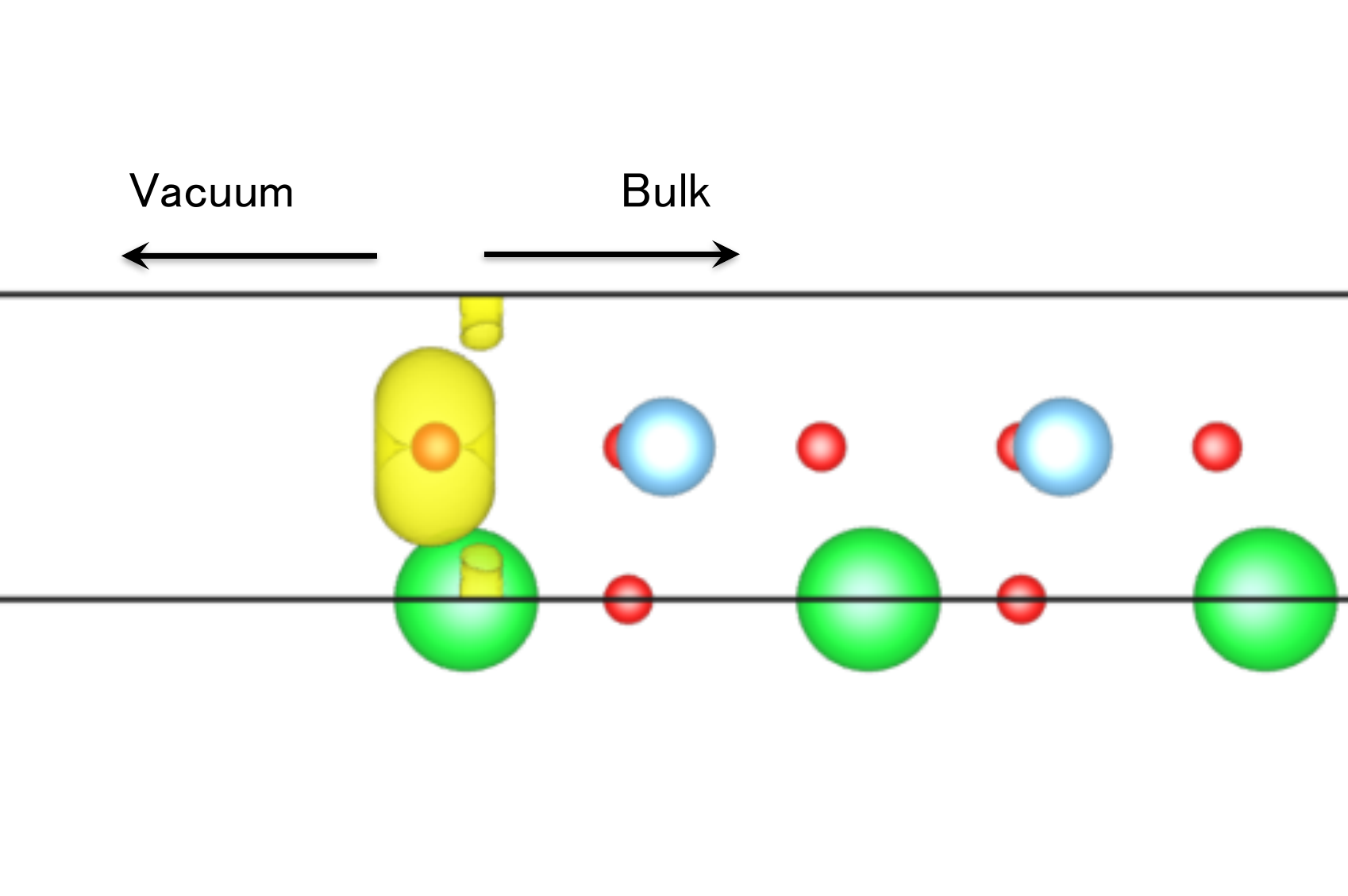}
  \end{minipage}
  \caption{\label{fig:figure3}
  (a) Surface-projected band structures for the p-type surface in the case of SrO terminations and the compressive strain of 7\%. The depth of the color of the dots denotes the weight of bands. SRSs also denote the SSs with the Rashba spin splitting for which the Rashba coefficient $\alpha_R$ = 155.7 meV$\cdot$\AA. The origin in energy is set to the Fermi level. The inset shows the enlarged view for SRSs. (b) Charge density distribution for SRSs. The green, blue, and red circles denote the Sr, Ti, O atoms, respectively.}
\end{figure}

\section{Conclusion}

We have performed first-principles density functional calculations for a compressive-strained SrTiO$_3$(001) ultra thin-film.
The 2DEG originating from the strain-induced polarization leads to the metallic SSs with large Rashba spin splittings.
The compressive strain and surface termination can control the spin splitting with $\alpha_R$ in the range of 10-100 meV$\cdot$\AA.
Our results suggest that 2DEG formed at ferroelectric oxide surfaces has possibilities for giant Rashba spin splittings applicable to spin-to-charge conversion.

\begin{acknowledgments}
  The authors thank H. Kotaka for the invaluable discussion about the analysis of surface-projected band structures.
  This work was supported by JSPS KAKENHI Grant Numbers JP16K04875, and JP17H05180.
  The computation was mainly carried out using supercomputers at ISSP, The University of Tokyo, computer facilities at RIIT, Kyushu University, and the K computer at AICS, RIKEN.
  This work was supported in part by MEXT as a social and scientific priority issue (Creation of new functional devices and high-performance materials to support next-generation industries) to be addressed using the post-K computer (Project ID: hp170269).
\end{acknowledgments}

\bibliographystyle{h-physrev5-1}
\bibliography{readcube_export}

\end{document}